\documentclass{kapproc}
\usepackage{procps}
\usepackage[dvips]{graphicx}

\upperandlowercase
\normallatexbib 

\begin{document}

\articletitle{Density Functional Calculations near Ferromagnetic Quantum
Critical Points}

\author{I.I. Mazin, D.J. Singh and A. Aguayo}
\altaffiltext{ }{Center for Computational Materials Science\\
Naval Research Laboratory\\
Washington, DC 20375}
\begin{abstract}
We discuss the application of the density functional theory in the
local density approximation (LDA) near a ferromagnetic quantum 
critical point.
The LDA fails to describe the critical fluctuations in this regime.
This provides a fingerprint of a materials near ferromagnetic
quantum critical points:
overestimation of the tendency to magnetism in the local density
approximation. This is in contrast to the typical, but not
universal, tendency of the LDA to underestimate the tendency to
magnetism in strongly Hubbard correlated materials.
We propose a method for correcting 
the local density calculations by including critical spin
fluctuations. This is based on (1) Landau expansion
for the free energy, evaluated within the LDA, (2) lowest order
expansion of the RPA susceptibility in LDA and (3) extraction of
the amplitude of the relevant (critical) fluctuations by 
applying the fluctuation-dissipation theorem to the difference
between a quantum-critical system and a reference system removed from the 
quantum critical point.
{ We illustrate some of the aspects of this by the}
 cases of Ni$_3$Al and Ni$_3$Ga, which are very
similar metals on opposite sides of a ferromagnetic quantum critical
point.
LDA calculations predict
that Ni$_3$Ga is the more magnetic system, but we find that due to
differences in the band structure, fluctuation effects are larger
in Ni$_3$Ga, explaining the fact that experimentally it is the less
magnetic of the two materials.
\end{abstract}


\section{Introduction}

Recent low temperature experiments on clean materials near ferromagnetic
quantum critical points (FQCP) have revealed a remarkable range of unusual
properties, including non-Fermi liquid scalings over a large phase space,
unusual transport, and novel quantum ground states, particularly coexisting
ferromagnetism and superconductivity in some materials. Although criticality
usually implies a certain universality, present experiments show
considerable material dependent aspects that are not well understood, \cite%
{laughlin} \emph{e.g.} the differences between UGe$_{2}$ and URhGe \cite%
{saxena,aoki} and ZrZn$_{2}$, \cite{pfleiderer} which both show coexisting
ferromagnetism and superconductivity but very different phase diagrams, in
contrast to MnSi, where very clean samples show no hint of superconductivity
around the QCP, {possibly because of the lack of the inversion symmetry}. \cite{pf2}

Moreover, by far not every magnetic material can be driven to a QCP by
pressure or by other means of supressing
ferromagnetism.
Typically, the transition becomes first
order as the Curie temperature, $T_C$ is depressed. If this happens
too far away from the fluctuation dominated regime, nothing interesting
is seen. Also, more pedestrian effects are often important. For example,
impurities or other defects can lead to scattering that smears out
the quantum critical region.

\section{The LDA Description Near a FQCP}

One of the fingerprints of a FQCP, maybe the most universal one, is a
substantial overestimation of the tendency to magnetism in conventional
density functional theory (DFT) calculations, such as within
the local density approximation (LDA).
Generally, approaches based on
density functional theory (DFT) are successful in accounting for material
dependence in cases where sufficiently accurate approximations exist.
Density functional theory is in principle an exact ground state theory. It
should, therefore, correctly describe the spin density of magnetic systems.
This is usually the case in actual state of the art density functional
calculations. However, common approximations to the exact density functional
theory, such as the LDA, may miss important physics and indeed fail to
describe some materials. A well know example is in strongly Hubbard
correlated systems, where the LDA treats the correlations in an orbitally
averaged mean field way and often
underestimates the tendency towards magnetism.

Overestimates of magnetic tendencies, especially in the LDA, are
considerably less common, the exceptions being materials near magnetic
quantum critical points (QCP);
here the error comes from neglect of low energy quantum spin fluctuations.
In particular, the LDA is parameterized based on the uniform electron gas at
densities typical for atoms and solids. However, the uniform electron gas at
these densities is stiff against magnetic degrees of freedom and far from
magnetic QCP's. Thus, although the LDA is exact for the uniform electron
gas, and therefore does include all fluctuation effects { there},
 its description of magnetic ground states in solids and
molecules is mean field like. This leads to problems such as the incorrect
description of singlet states in molecules with magnetic ions as well as
errors in solids when spin fluctuation effects beyond the mean field are
important. In solids near a QCP, the result is an overestimate of the
magnetic moments and tendency toward magnetism (\textit{i.e.} misplacement
of the position of the critical point) due to neglect of the quantum
critical fluctuations. \cite{yamada,millis} Examples include three types of
materials: paramagnets that are ferromagnetic in the LDA, ferromagnets where
the equilibrium magnetic moment is substantially overestimated in the
LDA, and
paramagnets where the paramagnetic susceptibility is substantially
overestimated.

\begin{table}
\caption{
Some materials near a FQCP that { we} have investigated
by LDA calculations.
Type 1
materials are ferromagnetic both in the calculations and in experiment;
magnetic moments in $\mu _{B}$ per formula unit are given. Type 2 are
ferromagnetic only in the calculations (calculated moments given, and
type 3 are paramagnetic (susceptibility in 10$^{-4}$emu/mol is given).
The references are to the LDA calculations.
}
\begin{tabular}{lllllll}
\hline
Material & ZrZn$_{2}$ & Ni$_{3}$Al &  Sc$_{3}$In & FeAl & Ni$_{3}$Ga \\
Type & 1 & 1 &  1 & 2 & 2   \\ 
Calc/Exp & 0.72/0.17 & 0.71/0.23 & 1.05/0.20 & 0.80 & 0.79 \\
Ref. & \cite{zrzn2} & \cite{ni3al} &  \cite{sc3in} & \cite{feal}
& \cite{ni3al} \\
\hline
\\
\hline
Material & Sr$_{3}$Ru$_{2}$O$_{7}$ & SrRhO$_{3}$ & Na$_{0.5}$CoO$_{2}$ & Pd \\ 
Type &  2 & 2 & 2 & 3 \\ 
Calc/Exp &  0.80 & 0.9  &  0.5 &  11.6/6.8 \\ 
Ref.   & \cite{sr-327} & \cite{SRhO} & \cite{nac2} & \cite{larson}\\
\hline
\end{tabular}
\label{materials}
\end{table}

We list examples of
materials in all three categories in Table \ref{materials}. At least two of these are
cases where a large deviation between the LDA and experimental
magnetic properties were noted, followed by 
{ transport measurements}
that suggest a nearby ferromagnetic quantum critical
point. In particular, in Sr$_3$Ru$_2$O$_7$, LDA calculations with
the experimental crystal structure found a sizeable moment, \cite{sr-327}
while
experimentally the material was known to be a paramagnetic metal.
Grigera and co-workers then showed that Sr$_3$Ru$_2$O$_7$ has
a metamagnetic quantum critical point at moderate field. \cite{grigera}
{ Pd metal provides another example: the calculated LDA magnetic
susceptibility is nearly twice larger that the experimental one. Correspondingly,
Nicklas $et$ $al$\cite{nicklas} found a FQCP in the Pd$_{1-x}$Ni$_x$ system at $x=0.026$,
where the transport properties become non-Fermi liquid.} 

We emphasize that substantial overestimates of the tendency of metals towards
ferromagnetism within the LDA is a rare occurance, and propose that it
be used as an indicator of critical fluctuations in a material. However,
for this to be an effective screen,
competing states, like antiferromagnetism need to be ruled out in each material.
An interesting case study is LiV$_2$O$_4$, which is a paramagnetic metal
and occurs in the cubic spinel structure. Remarkably, it was discovered
by Kondo and co-workers that this material behaves at low temperature like
a heavy fermion metal. \cite{kondo}
LDA calculations showed that the material is unstable against ferromagnetism
with a sizeable moment. \cite{eyert,anisimov,livo}
But calculations also show that the interactions are antiferromagnetic, and
as a result it is more unstable against antiferromagnetism, which however
is frustrated on the spinel lattice. While LiV$_2$O$_4$ may be near
an antiferromagnetic QCP, it is not a material near an FQCP.

\section{``Beyond-LDA'' Critical Fluctuations}

A popular way to add quantum or termal fluctuation to a mean-field type
theory is $via$ fluctuation corrections to Ginzburg-Landau expansion of the
free energy. For a detailed discussion we refer the reader to the book
of Moriya \cite{moriabook} and the review article of Shimizu
\cite{shimizu}. In short, one writes the free
energy (or the magnetic field) as a function of the ferromagnetic
magnetization, $%
M,$ 
\begin{eqnarray}
E_{static}(M) &=&a_{0}+\sum_{n\geq 1}\frac{1}{2n}a_{2n}M^{2n},  \label{Eexp}
\\
H_{static}(M) &=&\sum_{n\geq 1}a_{2n}M^{2n-1}  \label{Hexp}
\end{eqnarray}%
(obviously, $a_{2}$ gives the inverse spin susceptibility without
fluctuations), and then assume Gaussian zero-point fluctuations of an r.m.s.
magnitude $\xi $ for each of the $d$ components of the magnetic moment (for
a 3D isotropic material like Pd, $d=3).$ After averaging over the spin
fluctuations, one obtains a fluctuation-corrected functional. The general
expression can be written in the following compact form: 
\begin{eqnarray}
H(M) &=&\sum_{n\geq 1}\tilde{a}_{2n}M^{2n-1}  \nonumber \\
\tilde{a}_{2n} &=&\sum_{i\geq 0}C_{n+i-1}^{n-1}a_{2(n+i)}\xi ^{2i}\Pi
_{k=n}^{n+i-1}(1+\frac{2k}{d}).  \label{renorm}
\end{eqnarray}%
For instance,%
\begin{eqnarray}
\tilde{a}_{2} &=&a_{2}+\frac{5}{3}a_{4}\xi ^{2}+\frac{35}{9}a_{6}\xi ^{4}+%
\frac{35}{3}a_{8}\xi ^{6}...  \nonumber \\
\tilde{a}_{4} &=&a_{4}+\frac{14}{3}a_{6}\xi ^{2}+21a_{8}\xi ^{6}... 
\nonumber \\
&&...
\end{eqnarray}

The unrenormalized coefficients can be taken from fixed spin momen LDA
calculations, in which case $\xi $ becomes the amplitude of those
fluctuations only, which are not taken into account in LDA (as mentioned,
LDA includes some quantum fluctuation, specifically short-range fluctuations
present in the interacting uniform electron gas). In principle, one can
estimate $\xi $ from the fluctuation-dissipation theorem, which states that
(see, e.g., Refs. \cite{sol,kaul}) 
\begin{equation}
\xi ^{2}=\frac{4\hbar }{\Omega }\int d^{3}q\int \frac{d\omega }{2\pi }\frac{1%
}{2}\mathop{\rm Im}\chi (\mathbf{q},\omega ),  \label{mumu}
\end{equation}%
where $\chi (\mathbf{q},\omega )$ is the magnetic susceptibility and $\Omega 
$ is the Brillouin zone volume.
It is customary 
to approximate $\chi (\mathbf{q},\omega )$ by its small $q,$ small $\omega $ expansion
\cite{sol,kaul}:
\begin{eqnarray}
\chi _{0}(\mathbf{q},\omega ) &=&N(E_{F})-aq^{2}+ib\omega /q  \label{ab} \\
\chi ^{-1}(\mathbf{q},\omega ) &=&\chi _{0}^{-1}(\mathbf{q},\omega )- I,
\end{eqnarray}%

With the expansion (\ref{ab}) the integrations can be performed analytically, and the final 
result reads:
\begin{equation}
\xi ^{2}=\frac{bv_{F}^{2}N(E_{F})^{2}}{2a^{2}\Omega }[Q^{4}\ln
(1+Q^{-4})+\ln (1+Q^{4})].
\end{equation}
where 
$Q=q_{c}\sqrt{a/bv_{F}},$ and $q_{c}$ is the
cutoff parameter for momentum integration in Eqn. \ref{mumu}
(the frequency 
integration at a given $q$
is usually assumed to be cut off at $\omega=v_F q$).

To proceed along these lines one needs to find a way to
calculate the crucial parameters of the expansion (\ref{ab}).
 It was suggested by Moriya \cite{moriabook} that
these can be expressed as certain integrals over the Fermi surface, by
expanding the RPA expression for $\chi _{0}$. Below, we offer a derivation
equivalent to that of Moriya, but rendering the results in more computable
form. We start with the RPA expressions for the real and imaginary parts of $%
\chi _{0}:$ 
\begin{eqnarray}
\mathop{\rm Re}\chi _{0}(\mathbf{q,}0) &=&\sum_{\mathbf{k}}\left[ f(E_{%
\mathbf{k}})-f(E_{\mathbf{k+q}})\right] (E_{\mathbf{k+q}}-E_{\mathbf{k}%
})^{-1}  \label{Re} \\
\mathop{\rm Im}\chi _{0}(\mathbf{q,}\omega ) &=&\sum_{\mathbf{k}}[f(E_{%
\mathbf{k}})-f(E_{\mathbf{k+q}})]\delta (E_{\mathbf{k+q}}-E_{\mathbf{k}%
}-\omega ),  \label{Im}
\end{eqnarray}%
where $f(E)$ is the Fermi function, $-\frac{df(E)}{dE}=\delta (E-E_{F})$.
Expanding Eqn. \ref{Re} in $\Delta =E_{\mathbf{k+q}}-E_{\mathbf{k}}=\mathbf{v}%
_{\mathbf{k}}\mathbf{\cdot q+}\frac{1}{2}\sum_{\alpha \beta }\mu _{\mathbf{k}%
}^{\alpha \beta }q_{\alpha }q_{\beta }+...,$ we get to second order in $q$%
\begin{eqnarray*}
&&\mathop{\rm Re}\chi _{0}(\mathbf{q,0})=N(E_{F})
+\sum_{\mathbf{k}}[ 
\frac{1}{2}\left( \frac{d\delta (\varepsilon _{\mathbf{k}}-E_{F})}{dE_{F}}%
\right) (\mathbf{v}_{\mathbf{k}}\mathbf{\cdot q+}\frac 12\sum_{\alpha ,\beta
}\mu _{\mathbf{k}}^{\alpha \beta }q_{\alpha }q_{\beta })
\\&&
+\frac{1}{6}%
\left( \frac{d^{2}\delta (\varepsilon _{\mathbf{k}}-E_{F})}{dE_{F}^{2}}%
\right) (\mathbf{v}_{\mathbf{k}}\mathbf{\cdot q})^{2}] .
\end{eqnarray*}%
The odd powers of $\mathbf{v_{k}}$ cancel out and we get ($\alpha ,\beta
=x,y,z$)%
\begin{eqnarray*}
&&\mathop{\rm Re}\chi _{0}(\mathbf{q}) =
\\&&
N(E_{F})+\sum_{\alpha ,\beta }\frac{%
q_{\alpha }q_{\beta }}{4}\frac{d\left\langle N(E_{F})\mu ^{\alpha \beta
}\right\rangle }{dE_{F}}+\sum_{\alpha ,\beta }\frac{q_{\alpha }q_{\beta }}{6}%
\frac{d^{2}\left\langle N(E_{F})v_{\alpha }v_{\beta }\right\rangle }{%
dE_{F}^{2}} \\
&&=N(E_{F})+\frac{q^{2}}{4}\frac{d\left\langle N(E_{F})\mu
_{xx}\right\rangle }{dE_{F}}+\frac{q^{2}}{6}\frac{d^{2}\left\langle
N(E_{F})v_{x}^{2}\right\rangle }{dE_{F}^{2}},
\end{eqnarray*}%
%
%
where $v_{x}^{2}=v_{y}^{2}=v_{z}^{2},$ $\mu _{xx}=\mu _{yy}=\mu _{zz}.$ The
last equality assumes cubic symmetry; generalization to a lower symmetry is
trivial. Using the following relation,%
\[
\sum_{\mathbf{k}}\mathbf{\nabla }_{\mathbf{k}}F(\varepsilon _{\mathbf{k}%
})=\sum_{\mathbf{k}}\frac{dF(\varepsilon _{\mathbf{k}})}{d\varepsilon _{%
\mathbf{k}}}\mathbf{\nabla }_{\mathbf{k}}\cdot \varepsilon _{\mathbf{k}%
}=\sum_{\mathbf{k}}\frac{dF(\varepsilon _{\mathbf{k}})}{d\varepsilon _{%
\mathbf{k}}}\mathbf{v}_{\mathbf{k}}, 
\]%
one can prove that%
\begin{equation}
\frac{d^{2}\left\langle N(E_{F})v_{x}^{2}\right\rangle }{dE_{F}^{2}}=-\frac{%
d\left\langle N(E_{F})\mu _{xx}\right\rangle }{dE_{F}}.
\end{equation}%
Therefore 
\begin{equation}
\mathop{\rm Re}\chi _{0}(\mathbf{q})=N(E_{F})-\frac{q^{2}}{12}\frac{%
d^{2}\left\langle N(E_{F})v_{x}^{2}\right\rangle }{dE_{F}^{2}}  \label{rechi}
\end{equation}

Similarly, for Eqn. \ref{Im} one has 
\begin{equation}
\mathop{\rm Im}\chi _{0}(\mathbf{q,}\omega )=\sum_{\mathbf{k}}\left[ \left( -%
\frac{df(\varepsilon )}{d\varepsilon }\right) \omega \delta (\mathbf{v}_{%
\mathbf{k}}\mathbf{\cdot q}-\omega )\right]
\end{equation}

After averaging over the directions of $\mathbf{q,}$ this becomes, for small 
$\omega ,$%
\begin{eqnarray}
\mathop{\rm Im}\chi _{0}(q\mathbf{,}\omega ) &=&\frac{\omega }{2}\sum_{%
\mathbf{k}}\frac{\delta (\varepsilon _{\mathbf{k}})}{v_{\mathbf{k}}q}\theta
(v_{\mathbf{k}}q-\omega )=\frac{\omega }{2q}\left\langle
N(E_{F})v^{-1}\right\rangle  \nonumber \\
v &=&\sqrt{v_{x}^{2}+v_{y}^{2}+v_{z}^{2}}.  \label{imchi}
\end{eqnarray}%
Although in real materials
the Fermi velocity is obviously different along different
directions, it is still a reasonable approximation to introduce an average $%
v_{F}$. Then the above formulae reduce all parameters needed for estimating
the {\em r.m.s.} amplitude of the
spin fluctuations to four integrals over the Fermi
surface, specifically,
the density of states, $N(E_{F})$,
$a=\frac{1}{12}\frac{d^{2}\langle N(E_{F})v_{x}^{2}\rangle }{dE_{F}^{2}}$,
$b=\frac{1}{2}\left\langle N(E_{F})v^{-1}\right\rangle $ and
$v_{F}=\sqrt{3\frac{\left\langle N(E_{F})v_{x}^{2}\right\rangle }{N(E_{F})}}$.

The physical meaning of
these parameters is as follows. $a$ defines the rate at which the static
susceptibility $\chi (q,0)$ falls away from the zone center, {\em i.e.}
the extent to which the tendency to ferromagnetism is stronger than that to
antiferromagnetism. This translates into the phase space in the Brillouin
zone where the spin fluctuations are important. $b$ controls the dynamic
effects in spin susceptibility.

Note that the cutoff parameter $q_c$ 
remains the only undefined quantity in this
formalism. One obvious choice is $q_{c}=$ $\sqrt{N(E_{F})/a},$ because for
larger $q$ the approximation (\ref{ab}) gives unphysical negative values for
the static susceptibility. On the other hand, 
one may argue that $q_c$ should
reflect mainly the geometry of the Fermi surface 
and thus not depend on $a$ at all. We will come back to this issue 
later in this paper and will propose an approach that avoids using
$q_c$ whatsoever.

\section{Ni$_3$Al and Ni$_3$Ga}

Here we use the closely related compounds
Ni$_3$Al and Ni$_3$Ga to illustrate some of the above ideas.
Further details may be found in Ref. \cite{ni3al}.
These have the ideal
cubic Cu$_3$Au $cP4$ structure, with very similar lattice constants,
$a=3.568$ \AA ~and $a=3.576$ \AA, respectively,
and have been extensively studied by various
experimental techniques. Ni$_3$Al is a weak
itinerant ferromagnet,
$T_{c}$ = 41.5 K and magnetization, $M$=0.23 $\mu _{B}$/cell
(0.077 $\mu _{B}$/Ni atom) \cite{boer}
with a QCP under pressure
at $P_c$=8.1 GPa,
\cite{niklowitz}
while Ni$_3$Ga is a strongly renormalized paramagnet. \cite{hayden}
Further, it was recently reported that
Ni$_3$Al shows non-Fermi liquid transport over
a large range of $P$ and $T$ range down to very low $T$.
\cite{steiner}

\begin{figure}[tbp]
\vspace{.1in}
\centerline{
\includegraphics[height=4cm]{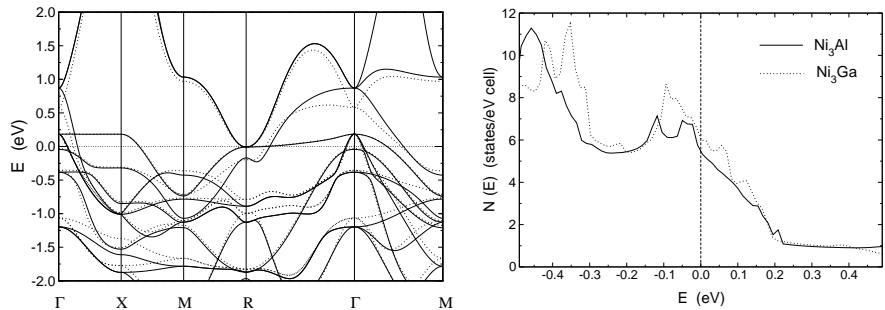}
\hskip 1.2cm 
\includegraphics[height=4cm]{nialga2z.eps}
}
\caption{Calculated LDA band structure (left)
and density of states (right) per f.u.~for
non-spin-polarized Ni$_{3}$Al (solid lines) and Ni$_{3}$Ga (dotted lines).
$E_F$ is at 0 eV. } 
\label{bands} \end{figure}

Previous LDA calculations showed that the
magnetic tendency of both materials is overestimated within the
LDA, and that Ni$_3$Ga is incorrectly predicted to be a ferromagnet.
\cite{buiting,moruzzi1,min,xu,guo,hsu}
Moreover, in the LDA the tendency to magnetism 
is stronger in Ni$_3$Ga than Ni$_3$Al,
{\em opposite to the experimental trend}.
This poses an additional challenge to any theory striving to 
describe the material dependent aspects of quantum criticality.
The two materials are expected to be very similar electronically
(the small difference between
the two is due to relativistic effects associated with Ga in Ni$_3$Ga).
Thus these two very similar metals offer
a very useful and sensitive benchmark for theoretical approaches. We use this
to test an approach based on the fluctuation dissipation theorem
applied to the LDA band structures with an ansatz for the cut-off
$q_c$. We find that this approach corrects the ordering of the
magnetic tendencies of the materials, and gives the right ground
states at ambient pressure as well as a reasonable value of $P_c$
for Ni$_3$Al.

\begin{figure}[tbp]
\vspace{0.3in}
\centerline{\includegraphics[width=2.5in]{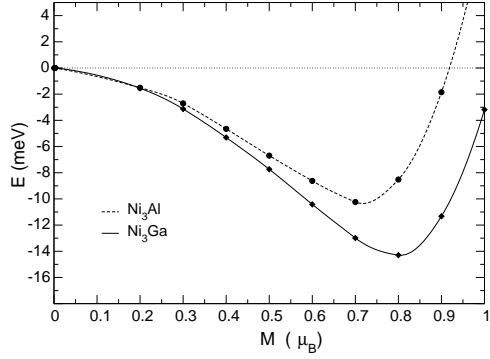}}
\caption{Energy {\em vs.} fixed spin moment for Ni$_3$Al and Ni$_3$Ga
at the experimental lattice parameters.
The energy zero is set to the non-spin-polarized value.}
\label{FSM} \end{figure}

\begin{figure}[tbp]
\centerline{\includegraphics[width=4in]{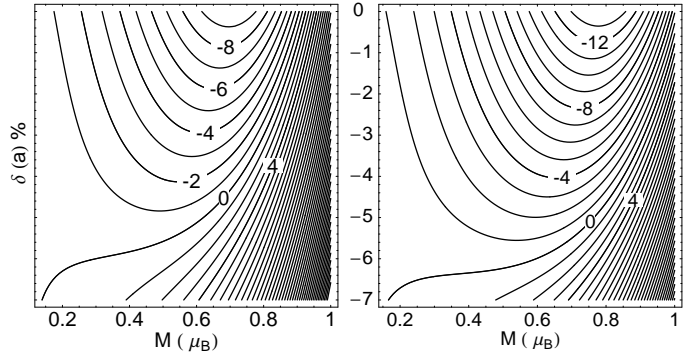}}
\caption{FSM calculations under hydrostatic pressures. Magnetic energy,
defined as the energy relative to the non-spin-polarized result
at the same volume,
as a function of the moment and linear compression.
Left and right panels correspond to Ni$_3$Al and
Ni$_3$Ga, respectively.}
\label{P}
\end{figure}

The LDA calculations were done using the
general potential linearized augmented planewave (LAPW) method
with local orbital extensions
\cite{singh-book,singh-lo,WIEN} as decribed in Ref. \cite{ni3al},
with the exchange-correlation
functional of Hedin and Lundqvist with the von Barth-Hedin
spin scaling \cite{hl,hl2}.
The LDA electronic structure is given in Fig. \ref{bands} and Table 2,
while results of fixed
spin moment calculations of the magnetic properties at the experimental
lattice parameters and under hydrostatic compression are given in
Figs. \ref{FSM} and \ref{P}.
The two compounds are
very similar in both electronic and magnetic properties, the main
apparent difference being the higher equilibrium moment of Ni$_3$Ga
(0.79 $\mu_B$/f.u. {\em vs.} 0.71 $\mu_B$/f.u.),
in agreement with other full potential calculations. \cite{guo,hsu}

The propensity
towards magnetism may be described in terms of the Stoner criterion, 
$IN(E_{F})$, where $I$ is the so-called Stoner parameter, which derives
from
Hund's rule coupling on the atoms. For finite magnetizations,
the so-called extended Stoner model \cite{Krasko}, states
that, to the second order in the spin density, the magnetic stabilization
energy is given by
\begin{equation}
\Delta E={M^{2}}[\int_0^Mm~dm/2\tilde{N}(m)-I/4],  \label{eStoner}
\end{equation}
where $\tilde{N}(M)$ is the density of states averaged over the exchange
splitting corresponding to the magnetization $M.$
Fitting the fixed spin moment results to this expression,
we find $I_{Al}=0.385$ eV and $I_{Ga}=0.363$ eV. These gives
$IN(E_F)=$1.21
and $IN(E_F)$ = 1.25 for Ni$_{3}$Al and Ni$_{3}$Ga, respectively.
Both numbers are larger than unity,
corresponding to a ferromagnetic instability,
and the value for Ni$_{3}$Ga is larger than that for Ni$_{3}$Al.
Importantly, the difference comes from the density of states, since
$I_{Al}>I_{Ga}$.
In both compounds, magnetism is suppressed by
compression, with an LDA critical point at a value $\delta a/a \sim$
-0.05 -- -0.06. In Ni$_3$Al, the critical point at $\delta a/a$ =-0.058
corresponds to the pressure of $P_c=$50 GPa, \cite{Pnote} which is
much higher than the experimental value.
It is interesting that, as in ZrZn$_2$ \cite{zrzn2},
the exchange splitting is very strongly {\bf k}-dependent;
for instance, in Ni$_3$Al at some points it is as small as 40 meV/$\mu_B$
near the Fermi level, while at the others (of pure Ni d character) it is
close to 220 meV/$\mu_B$.

Notwithstanding
the general similarity of the two compounds,
there is one important difference near the Fermi level,
specifically, the light band crossing the Fermi level in the middle
of the $\Gamma $-M or $\Gamma $-X directions is
steeper in Ni$_{3}$Al (Fig. \ref{bands}). 
This,
in turn, leads to smaller density of states. This comes from
a different position of the top of this band at the $%
\Gamma $ point, 0.56 eV in Ni$_{3}$Ga and 0.85 eV in Ni$_{3}$Al. The
corresponding electronic state is a mixture of Ni $p$ and Al (Ga) $p$
states, and is the only state near the Fermi level with substantial Al (Ga)
content. Due to relativistic effects, the Ga $p$ level is lower than the Al $%
p$ level and this leads to the difference in the position of the
corresponding hybridized state. Note that this is a purely scalar relativistic
effect. Including spin orbit does not produce any further
discernible difference.

\vspace{0.15in}

Returning to magnetism, the
fixed spin moment calculations provide the energy $E$ as a function of the
magnetization $M$
(Fig. \ref{FSM}).
One can write a Landau expansion for $E(M)$ as in Eqn. \ref{Eexp},
which may then be treated as a mean field expression adding the
effects of spin fluctuations. \cite{shimizu}

Treating this as a mean field expression and adding the effects of spin
fluctuations \cite{shimizu} leads to renormalization of the expansion
coefficients.
The renormalized coefficients $\tilde{a}_{i}$ are
written as power series in the averaged square of the magnetic moment
fluctuations beyond the LDA, 
$\xi ^{2}$ as in Eqn. \ref{renorm}.
$\xi $ may then be estimated by requiring that the
corrected Landau functional reproduces the experimental magnetic
moment (for Ni$_{3}$Al) or experimental magnetic susceptibility (for Ni$_{3}$%
Ga). The ``experimental" $\xi $'s
obtained in this manner are
are 0.47 and 0.55, respectively, which implies that spin fluctuation
effects must be stronger in Ni$_{3}$Ga than in Ni$_{3}$Al.

\begin{table}
\caption{Magnetic energy (see text), magnetic moment in $\protect\mu _{B}$/cell
and $N(E_{F})$ in eV$^{-1}$ for Ni$_3$Al and Ni$_3$Ga
on a per spin per formula unit basis.}
\begin{tabular}{ccccc}
  & $|\Delta E|$ (meV)&$M$ (calc.)&$M$(expt.)&  $N(E_F)$ \\
\hline
Ni$_3$Al & 10.3 & 0.71  & 0.23  &  3.2  \\
Ni$_3$Ga & 14.3 & 0.79  & 0.00  &  3.4  \\
\end{tabular} 
\label{tab:table1} \end{table}

A link can now be made
between this fact and the electronic structures,
using the formalism outlined in the previous section.
As discussed, the cutoff parameter $q_{c}$ is the least well defined quantity in this
formalism. 
Furthermore, the fermiology of these compounds is very complicated:
in the paramagnetic state, there are four Fermi surfaces, two small and two
large (one open and one closed). In this situation, it is hardly
possible to justify any simple prescription for $q_c$. Therefore, we
chose a different route: we assume that $q_c$ is the {\it same} for
both materials, and choose a number which yields a good description of
both the equilibrium moment in Ni$_3$Al and the paramagnetic susceptibility
in Ni$_3$Ga, $q_c=0.382$ $a_0^{-1}$. Note that this is larger that the diameters
of the small Fermi surfaces but smaller than the radius of the Brillouin
zone, $\approx 0.5$ $a_0^{-1}$.

To calculate the above quantities, especially $a,$ we need accurate values
of the velocities on a fine mesh. Numerical differentiation of energies
within the tetrahedron method proved to be too noisy. Therefore we use the
velocities obtained analytically as matrix elements of the momentum
operator, computed within the \textit{optic} program of the WIEN package. A
bootstrap method, \cite{boot}
as described in Ref. \cite{larson}, was used to obtain
stable values for $a,b$. We found for Ni$_{3}$Al,
using as the energy unit Ry, the length unit Bohr, and
the velocity unit Ry$\cdot $Bohr,
$a=230$, $b=210$, $v_{F}=0.20$,  and $\xi =0.445$ $\mu_B$.
For Ni$_{3}$Ga $a=140$, $b=270$, $v_{F}=0.19$,  and $%
\xi =0.556$ $\mu_B$. Using the resulting values of $\xi $
each compound we obtain a magnetic moment of $M=0.3$ $\mu_B $/cell for Ni$%
_{3}$Al and a paramagnetic state with the renormalized susceptibility $\chi
(0,0)=1/\tilde{a}_{2}=6.8\times10^{-5}$ emu/g for Ni$_{3}$Ga, thus correcting
the incorrect ordering of the magnetic tendencies of these two compounds
and reproducing extremely well the experimental numbers of $M=0.23$ $\mu_B$, $\chi
(0,0)=6.7\times 10^{-5}$ emu/g, respectively.
This qualitative behavior is due to
the different coefficient $a,$ {\em i.e.}, different $q$
dependencies of $\chi _{0}(q,0)$ at small $q$,
which relates to the phase space available for soft fluctuations.

Now we turn to the pressure dependence. The above results imply that
beyond-LDA fluctuations are already larger than the moments themselves
at $P=0$. In this regime, we may assume that the size of the
beyond-LDA fluctuations is only weakly pressure dependent.
Then we can apply Eqn.
\ref{renorm} to the data shown in Fig. \ref{P}
using $\xi=0.47$ as needed to match the $P=0$ value of $M$.
This yields a value $P_c$=10 GPa in quite good agreement with
the experimental value, $P_c$=8.1 GPa.
\cite{niklowitz}

\section{Towards a Fully First Principles Theory}

The results for Ni$_3$Al and Ni$_3$Ga, discussed above, and in Ref.
\cite{ni3al}, show that an approach based on correction of the
LDA using the fluctuation dissipation theorem has promise. However,
the results hinge on an unknown cut-off, which serves the purpose
of including fluctuations that are associated with the FQCP and
are not included in the LDA, from those that are included in the LDA.
While it is apparently possible to obtain useful results using
reasonable ansatz for this cut-off, it would be much better to have
a truly first principles theory, with no parameters. In order to
construct such a theory, one should find a way of solving the double
counting problem, {\em i.e} including in the correction only those
fluctuations that are not already taken into acount at the LDA level.
This amounts to subtracting from Eqn. \ref{mumu} the fluctuations already
included in the LDA.
Since the LDA is known to work well for materials far from an FQCP,
this means that the correction should be zero or close to it for
the most materials.

We suggest that a consistent way to accomplish this is by introducing a
``reference" susceptibility $\chi ^{ref}(%
\mathbf{q},\omega )$ and subtracting it from $\chi (\mathbf{q},\omega ):$

\begin{equation}
\xi ^{2}=\frac{4\hbar }{\Omega }\int d^{3}q\int \frac{d\omega }{2\pi }\frac{1%
}{2}\mathop{\rm Im}[\chi (\mathbf{q},\omega )-\chi ^{ref}(\mathbf{q},\omega
)],  \label{mumuref}
\end{equation}

We shall use the same expansion\ref{ab} for both $\chi (\mathbf{q},\omega )$
and $\chi ^{ref}(\mathbf{q},\omega)$, to derive equivalent
expansions 
\begin{equation}
\chi ^{-1}(\mathbf{q},\omega )=\chi _{0}^{-1}(0,0)-I+Aq^{2}-iB\omega /q,
\label{expchi}
\end{equation}%
where $\chi _{0}^{-1}(0,0)=1/N(E_{F})$ (density of states per spin) is the
bare (noninteracting) static uniform susceptibility, and the Stoner
parameter $I$ is only weakly dependent on \textbf{q} and $\omega $. 
Note that $A=a/N^{2},$ $B=b/N^{2},$ where $a$ and $b$ are the coefficients
introduced in Eq.\ref{ab}. We also introduce a notation, $\Delta
=N(E_{F})^{-1}-I.$ As long the same functional form (\ref{expchi}) is used
for $\chi (\mathbf{q},\omega )$ and $\chi ^{ref}(\mathbf{q},\omega ),$ the
condition for the convergence of the integral (\ref{mumuref}) is that the
coefficients $A$ and $B,$ controlling the short-range and high frequency
fluctuations are the same. Of course, the parameter $\Delta$, defining the
proximity to the QCP, is different in the reference system,
which like the uniform electron gas upon which the LDA is based, should
be far from any QCP
(let us call $\Delta$ for the reference system $\Delta _{0}$).

To calculate the integral ((\ref{mumuref}), we write it in the
following form:%
\begin{equation}
\xi ^{2}=\frac{4\hbar }{\Omega }\int d^{3}q\int \frac{d\omega }{2\pi }\frac{1%
}{2}\mathop{\rm Im}[\chi (\Delta ,\mathbf{q},\omega )-\chi (\Delta
_{0},\mathbf{q},\omega )].
\end{equation}%
For instance, $\chi (0,\mathbf{q},\omega )$ represents the susceptibility
right at the FQCP. This diverges for $\mathbf{q}=0$, $\omega=0$.
The derivation then proceeds as follows:

\begin{equation}
\int^{\omega _{c}}d\omega \mathop{\rm Im} [\chi (\Delta ,q,\omega )]=\frac{q}{2B}%
\ln [\frac{(\Delta +Aq^{2})^{2}+B^{2}\omega _{c}^{2}/q^{2}}{(\Delta
+Aq^{2})^{2}}].
\end{equation}%
Where we introduce the Landau cutoff frequency, $\omega _{c}=vq$ (here $v$
is an average Fermi velocity) and the notation $\beta =Bv$. We will also need
the following function:%
\begin{eqnarray}
F(\Delta ,\beta ,x) &=&\int x^{3}dx\ln [(\Delta +x^{2})^{2}+\beta ^{2}] \nonumber\\
&=&\frac{(\Delta +x^{2})^{2}+\beta ^{2}}{4}\{\ln [(\Delta +x^{2})^{2}+\beta
^{2}]-1\} \nonumber \\
&-&\frac{\Delta (\Delta +x^{2})}{2}\{\ln [(\Delta +x^{2})^{2}+\beta
^{2}]-2\}+\beta \Delta \tan ^{-1}\frac{\beta }{\Delta +x^{2}\ }\nonumber
\end{eqnarray}%
Now 
\begin{eqnarray*}
\xi ^{2} &=&\frac{2}{\Omega A^{2}B}\lim_{Q\rightarrow \infty }[F(\Delta
,\beta ,Q)-F(\Delta _{0},\beta ,Q) \\
&&-F(\Delta ,0,Q)+F(\Delta _{0},0,Q)-F(\Delta ,\beta ,0) \\
&&+F(\Delta _{0},\beta ,0)+F(\Delta ,0,0)-F(\Delta _{0},0,0)].
\end{eqnarray*}%
This is particularly easy to evaluate at $\Delta =0$. The result is%
\begin{eqnarray*}
&&\Xi ^{2}(\Delta _{0}) =\\
&&\frac{2}{\Omega A^{2}B}[\Delta _{0}\beta (\frac{\pi 
}{2}-\tan ^{-1}\frac{\Delta _{0}}{\beta })+\frac{\beta ^{2}-\Delta _{0}^{2}}{%
4}\ln \frac{\Delta _{0}^{2}+\beta ^{2}}{\beta ^{2}}+\frac{\Delta _{0}^{2}}{4}%
\ln \frac{\Delta _{0}^{2}}{\beta ^{2}}]  \label{xi} \\
&&\Xi ^{2}(S_{0}) =\frac{N^{2}bv_{F}^{2}}{2\Omega a^{2}}[4S_{0}\tan
^{-1}(S_{0}^{-1})+\ln (1+S_{0}^{2})-S_{0}^{2}\ln (1+S_{0}^{-2})],
\label{xi2}
\end{eqnarray*}%
where $S_{0}=\Delta _{0}N^{2}/bv_{F}$. Obviously, for arbitrary $\Delta $
the answer is simply%
\begin{equation}
\xi ^{2}=\Xi ^{2}(S_{0})-\Xi ^{2}(S).
\end{equation}%
\newline
Given that usually the reason for a quantum criticality is a large density of
states, it makes sense to take the Stoner parameter for the reference system
the same as for the system in question.
The point is that the density of states is a highly non-local parameter
(note that it involves a delta function integral in energy), which can
hardly be discerned from local information about the charge density,
while the Stoner parameter is a very local quantity associated with
the exchange-correlation potential.
The difference between $\Delta $ and 
$\Delta _{0}$ then comes from the difference between $N=N(E_{F})$ and the
density of states, $N_{0},$ of the reference system.

One may think about several different ways for choosing $N_{0}.$ One may be
to take average $N(E)$ over the width of the valence band, $N_{0}=n/t,$
where $n$ is the total number of states in the band and $t$ is its width.
One can also think about the density of states of the uniform electron gas
with the same Stoner parameter. There may be other, more sophisticated
prescriptions. Probably, the most practical approach will be found after
several trial and error tests with real materials.

\section{Summary and Open Questions}

The failure of the usual approximations to density functional theory,
for example, the LDA,
to describe the magnetic properties of materials near ferromagnetic
quantum critical points is associated with renormalization due to
critical fluctuations. It is pointed out that since such fluctuations
are invariably antagonistic to ferromagnetic ordering, deviations
between experiment and LDA calculations in which the LDA is overly
ferromagnetic can be a useful screen for materials near FQCPs.
These errors in the LDA can be corrected using a phenomenalogical
Landau function approach with the fluctuation amplitude as a parameter.
However, there is hope that this parameter can be obtained from the
electronic structure via the fluctuation dissipation theorem and
a suitable reference system. The key remaining challenges in our
view are to define the reference system to be used, and to use
calculations to determine the usefulness of this approach for
real materials near a critical point.

\begin{acknowledgments}
We are grateful for helpful conversations with S.V. Halilov,
G. Lonzarich and S. Saxena.
Work at the Naval Research Laboratory is supported by the Office of
Naval Research.
\end{acknowledgments}


\begin{chapthebibliography}{99}

\bibitem{laughlin} R.B. Laughlin, G.G. Lonzarich, P. Monthoux and D. Pines,
Adv. Phys. \textbf{50}, 361 (2001).

\bibitem{saxena} S.S. Saxena, P. Agarwal, K. Ahilan, F.M. Grosche, R.K.W.
Haselwimmer, M.J. Steiner, E. Pugh, I.R. Walker, S.R. Julian, P. Monthoux,
G.G. Lonzarich, A. Huxley, I. Sheikin, D. Braithwaite, and J. Flouquet,
Nature \textbf{406}, 587 (2000).

\bibitem{aoki} D. Aoki, A. Huxley, E. Ressouche, D. Braithwaite, J.
Flouquet, J.P. Brison, E. Lhotel, and C. Paulsen, Nature \textbf{413}, 613
(2001).

\bibitem{pfleiderer} C. Pfleiderer, M. Uhlarz, S.M. Hayden, R. Vollmer, H.
von Lohneysen, N.R. Bernhoeft, and G.G. Lonzarich, Nature \textbf{412}, 58
(2001).

\bibitem{pf2} C.P. Pfleiderer, S.R. Julian and G.G. Lonzarich, Nature 
\textbf{414}, 427 (2001).

\bibitem{yamada} H. Yamada, K. Fukamichi and T. Goto, Phys. Rev. B \textbf{65%
}, 024413 (2001).

\bibitem{millis} A.J. Millis, A.J. Schofield, G.G. Lonzarich and S.A.
Grigera, Phys. Rev. Lett. \textbf{88}, 217204 (2002).

\bibitem{sr-327} D.J. Singh and I.I. Mazin, Phys. Rev. B \textbf{63}, 165101
(2001).

\bibitem{grigera}
S.A. Grigera, R.S. Perry, A.J. Schofield, M. Chiao, S.R. Julian, G.G.
Lonzarich, S.I. Ikeda, Y. Maeno, A.J. Millis, and A.P. Mackenzie,
Science \textbf{294}, 329 (2001).

\bibitem{nac2}
D.J. Singh, Phys. Rev. B {\bf 61}, 13397 (2000);
D.J. Singh, Phys. Rev. B \textbf{68}, 020503 (2003).


\bibitem{nicklas} M. Nicklas, M. Brando, G. Knebel, F. Mayr, W. Trinkl, and A. Loidl, 
 Phys. Rev. Lett. {\bf 82}, 4268 (1999).

\bibitem{kondo}
S. Kondo, D.C. Johnston, C.A. Swenson, F. Borsa, A.V. Mahajan, L.L. Miller,
T. Gu, A.I. Goldman, M.B. Maple, D.A. Gajewski, E.J. Freeman, N.R. Dilley,
R.P. Dickey, J. Merrin, K. Kojima, G.M. Luke, Y.J. Uemura,
O. Chmaissem and J.D. Jorgensen, Phys. Rev. Lett. {\bf 78}, 3729 (1997).

\bibitem{livo} D.J. Singh, P. Blaha, K. Schwarz, and I. I. Mazin, Phys.
Rev. B \textbf{60}, 16359 (1999).

\bibitem{eyert}
V. Eyert, K.H. Hock, S. Horn, A. Loidl and P.S. Riseborough,
Europhysics Lett. {\bf 46}, 762 (1999).

\bibitem{anisimov}
V.I. Anisimov, M.A. Korotin, M. Z\"olfl, T. Pruschke, K. Le Hur and T.M. Rice,
Phys. Rev. Lett. {\bf 83}, 364 (1999).

\bibitem{zrzn2}
D.J. Singh and I.I. Mazin,
Phys. Rev. B {\bf 69}, 020402 (2004).

\bibitem{ni3al}
A. Aguayo, I.I. Mazin, and D.J. Singh, cond-mat/0310629.

\bibitem{sc3in} A. Aguayo and D.J. Singh, Phys. Rev. B \textbf{66}, 020401
(2002).

\bibitem{feal} A.G. Petukhov, I.I.Mazin, L. Chioncel and A. I. Lichtenstein,
Phys. Rev. B \textbf{67}, 153106 (2003).

\bibitem{SRhO}
D.J. Singh, Phys. Rev. B {\bf 67}, 054507 (2003).

\bibitem{larson} P. Larson, I.I. Mazin and D.J. Singh, cond-mat/0305407
(Phys. Rev. B, January 2004, in press).

\bibitem{moriabook} T. Moriya, \textit{Spin fluctuations in itinerant
electron magnetism} (Berlin, Springer, 1985).

\bibitem{shimizu}
M. Shimizu, Rep. Prog. Phys. {\bf 44}, 329 (1981).

\bibitem{sol} A.Z. Solontsov and D. Wagner, Phys. Rev. \textbf{B51}, 12410
(1995).

\bibitem{kaul} S.N. Kaul, J. Phys. Cond. Mat. \textbf{11}, 7597 (1999).

\bibitem{boer} F.R. de Boer,
C.J. Schinkel, J. Biesterbos, and S. Proost,
J. Appl. Phys. \textbf{40}, 1049 (1969).

\bibitem{niklowitz}
P.G. Niklowitz, F. Beckers, N. Bernhoeft, D. Braithwaite,
G. Knebel, B. Salce, J. Thomasson, J. Floquet and G.G. Lonzarich
(unpublished); presented at Conference on Quantum Complexities in
Condensed Matter, 2003.

\bibitem{hayden} S.M. Hayden, G.G. Lonzarich and H.L. Skriver,
Phys. Rev. B \textbf{33}, 4977
(1986).

\bibitem{steiner}
M.J. Steiner, F. Beckers, P.G. Nicklowitz and G.G. Lonzarich,
Physica B {\bf 329}, 1079 (2003).

\bibitem{buiting} J.J. Buiting, J. Kl\"ubler, and F.M. Mueller,
J. Phys. F {\bf 39} L179 (1983).

\bibitem{moruzzi1} V.L. Moruzzi and P.M. Marcus, Phys. Rev. B,
{\bf 42}, 5539  (1990).

\bibitem{min} B.I. Min, A.J. Freeman, and H.J.F. Jansen, Phys. Rev. B
\textbf{37}, 6757 (1988).

\bibitem{xu} J.H. Xu,
B.I. Min, A.J. Freeman, and T. Oguchi,
Phys. Rev. B
\textbf{41}, 5010 (1990).

\bibitem{guo} G.Y Guo, Y.K. Wang, Li-Shing Hsu, J. Magn. Magn. Mater.
\textbf{239}, 91 (2002).

\bibitem{hsu}
L.-S. Hsu, Y.-K. Wang and G.Y. Guo, J. Appl. Phys. {\bf 92}, 1419 (2002).

\bibitem{singh-book}
D.J. Singh, {\em Planewaves Pseudopotentials and the LAPW Method}
(Kluwer Academic, Boston, 1994).

\bibitem{singh-lo} D. Singh, Phys. Rev. B
{\bf 43}, 6388 (1991).

\bibitem{wei} S.H. Wei and H. Krakauer, Phys. Rev. Lett.  {\bf 55}, 1200 (1985).

\bibitem{WIEN} P. Blaha, K. Schwarz G.K.H. Madsen, D. Kvasnicka, and J.
Luitz, \textit{WIEN2K, An Augmented Plane Wave + Local Orbitals Program for
for Calculating Crystal Properties} (K. Schwarz, Techn. Universitat Wien,
Austria, 2001), ISBN 3-9501031-1-2.

\bibitem{hl} L. Hedin and B. Lundqvist, J. Phys. C, {\bf 4}, 2064 (1971).
\bibitem{hl2}U. von Barth and L. Hedin, J. Phys. C {\bf 5}, 1629 (1972).

\bibitem{Krasko} G.L. Krasko, Phys. Rev. B, {\bf 36} 8565 (1987).

\bibitem{Pnote} To compute pressure, we used
$P=B/B'[(V/V_0)^{B'}-1]$, where $V/V_0$ is the volume compression,
$B$ and $B'$ are
the bulk modulus and its derivative. We used the experimental bulk
modulus of Ni$_3$Al, $B$=174 GPa. \cite{wallow}
For $B'$ we used the calculated value $B'$=5.2.

\bibitem{boot}B. Efron and R.J. Tibshirani, {\em
An Introduction to the Bootstrap} (Chapmann and Hall, New York, 1993).

\bibitem{schinkel} C.J. Schinkel, F.R. de Boer, and B. de Hon, J.
Phys. F \textbf{3}, 1463 (1973).

\bibitem{wallow}
F. Wallow, G. Neite, W. Schroer and E. Nembach,
Phys. Status Solidi A {\bf 99}, 483 (1987).

\end{chapthebibliography}
\end{document}